%% file: root.tex
\documentclass[letterpaper, 10 pt, conference]{ieeeconf}  

\IEEEoverridecommandlockouts       
\usepackage[utf8]{inputenc}
\usepackage{mathtools}
\usepackage{amssymb,amsmath}
\usepackage{ifthen}
\usepackage{color}
\usepackage{xspace}
\usepackage{lipsum}
\usepackage{dblfloatfix}
\PassOptionsToPackage{hyphens}{url}\usepackage{hyperref}
\usepackage{pgfplots,booktabs}
\usepackage{tikz}
\usetikzlibrary{shapes,decorations.pathmorphing,patterns}

\newboolean{showcomments}
\setboolean{showcomments}{true}
\ifthenelse{\boolean{showcomments}}
{ \newcommand{\mynote}[3]{
    \fbox{\bfseries\sffamily\scriptsize#1}
    {\small$\blacktriangleright$\textsf{\emph{\color{#3}{#2}}}$\blacktriangleleft$}}}
{ \newcommand{\mynote}[3]{}}

\title{\LARGE \bf
Secure Stream Processing for Medical Data
}

\author{Carlos Segarra$^{1}$, Enric Muntan\'e$^{1}$, Mathieu Lemay$^{1}$, Valerio Schiavoni$^{2}$, and Ricard Delgado-Gonzalo$^{1}$
\thanks{$^{1}$Carlos Segarra, Enric Muntan\'e, Mathieu Lemay, and Ricard Delgado-Gonzalo are with the Swiss Center for Electronics and Microtechnology (CSEM), Neuch\^atel, Switzerland
        {\tt\small carlos.segarra@csem.ch}}%
\thanks{$^{2}$Valerio Schiavoni is with the Universit\'e de Neuch\^atel, Neuch\^atel, Switzerland
        {\tt\small valerio.schiavoni@unine.ch}}%
}

\begin{document}

\maketitle
\thispagestyle{empty}
\pagestyle{empty}

\begin{abstract}
Medical data belongs to whom it produces it. In an increasing manner, this data is usually processed in unauthorized third-party clouds that should never have the opportunity to access it. Moreover, recent data protection regulations (\textit{e.g.}, GDPR) pave the way towards the development of privacy-preserving processing techniques. In this paper, we present a proof of concept of a streaming IoT architecture that securely processes cardiac data in the cloud combining trusted hardware and Spark. The additional security guarantees come with no changes to the application's code in the server. We tested the system with a database containing ECGs from wearable devices comprised of 8 healthy males performing a standardized range of in-lab physical activities (\textit{e.g.}, run, walk, bike). We show that, when compared with standard \textsc{Spark Streaming}, the addition of privacy comes at the cost of doubling the execution time.
\end{abstract}

\input{sections/introduction}
\input{sections/motivation.tex}
\input{sections/architecture.tex}
\input{sections/materials.tex}
\input{sections/evaluation.tex}
\input{sections/conclusion.tex}

{\footnotesize
    \bibliographystyle{unsrt}
    \bibliography{references}{}
}

\end{document}

%% file: sections/introduction.tex
\section{INTRODUCTION} \label{sec:introduction}

Personalized health and medicine has the potential of being the next revolution in healthcare. It is also referred as P4 medicine (Predictive, Preventive, Personalized, and Participatory), and it provides the opportunity to benefit from more targeted and effective diagnoses and treatments~\cite{Cumming2014}. To implement this, larger amounts of data and complex processing pipelines are gradually being deployed, what generally leads to offloading computation to third-party cloud providers. When the data-in-motion are vital signs, protecting user's privacy becomes a topic of crucial importance. Furthermore, recent data protection regulations (\textit{e.g.}, GDPR~\cite{Voigt2017,gdpr}) stress the importance of protecting sensitive information against malicious attackers or untrusted cloud providers.

The current state of the art for privacy-preserving computation falls in two big categories: homomorphic encryption or trusted hardware. Homomorphic encryption (HE) is a cryptographic scheme that allows evaluating algorithms over encrypted data without having to decrypt it~\cite{Gentry2009}. In spite of reducing the trusted computing base (TCB) to zero on the remote side, HE frameworks are currently prohibitively slow~\cite{Goettel2018}. On the other hand, we have Trusted Execution Environments (TEE). A TEE is an isolated area of a processor that grants \emph{confidentiality} and \emph{integrity} to the information therein contained. TEEs are nowadays available in commodity CPUs, namely \textsc{ARM TrustZone} and \textsc{Intel\textregistered\xspace SGX}. The \textsc{Intel\textregistered\xspace Software Guard eXtensions (SGX)} is a set of instructions and memory access changes added to the Intel\textregistered\xspace architecture. These instructions enable applications to create hardware-protected areas in their application memory address space called enclaves~\cite{McKeen2013}.

In this paper we present a real-time privacy-preserving streaming platform for performing cloud computing on live cardiac data that runs unmodified Spark applications in a distributed environment, performs critical parts of the computation inside enclaves, and provides end-to-end protection for users' data. Without any modification to their applications' source code, potential users would instantly benefit of privacy-preserving computing relying on Intel SGX technology with limited impact on the performance. The main contributions of this paper are: (i) a proof of concept for a privacy-preserving streaming platform for medical data, and (ii) a study on the overhead introduced by privacy-preserving processing techniques.

The paper is organized as follows. In Section~\ref{sec:introduction}, we introduce the problem we are facing and summarize the proposed solution. In Section~\ref{sec:motivation}, we describe our use case. In Section~\ref{sec:architecture}, we describe the architecture of the solution. Its components and materials are introduced in Section~\ref{sec:materials}. The experiments and results obtained are presented in Section~\ref{sec:evaluation}. Lastly, in Section~\ref{sec:conclusion}, we expose our main conclusions and propose further lines of research.

%% file: sections/motivation.tex
\section{MOTIVATION} \label{sec:motivation}

\begin{figure}[b!]
    \centering
    \input{resources/figs/ecg-hrv.tex}
    \caption{Schematic representation of an ECG signal showing three normal beats. A normal electrocardiogram can be broken down in three waves: a \textit{P wave} corresponding to the depolarization of the atria, a \textit{QRS complex} corresponding to the depolarization of the ventricles and a \textit{T wave} corresponding to the repolarization of the ventricle~\cite{Lilly2001}. From an ECG the sensor extracts and streams the R-peaks' timestamp and the time elapsed between them. \label{fig:ecg-hrv}}
\end{figure}
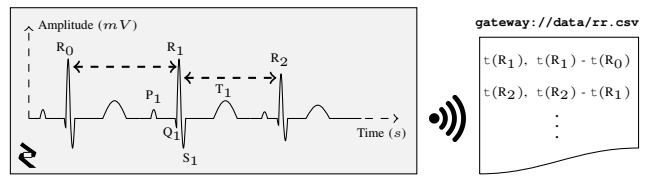

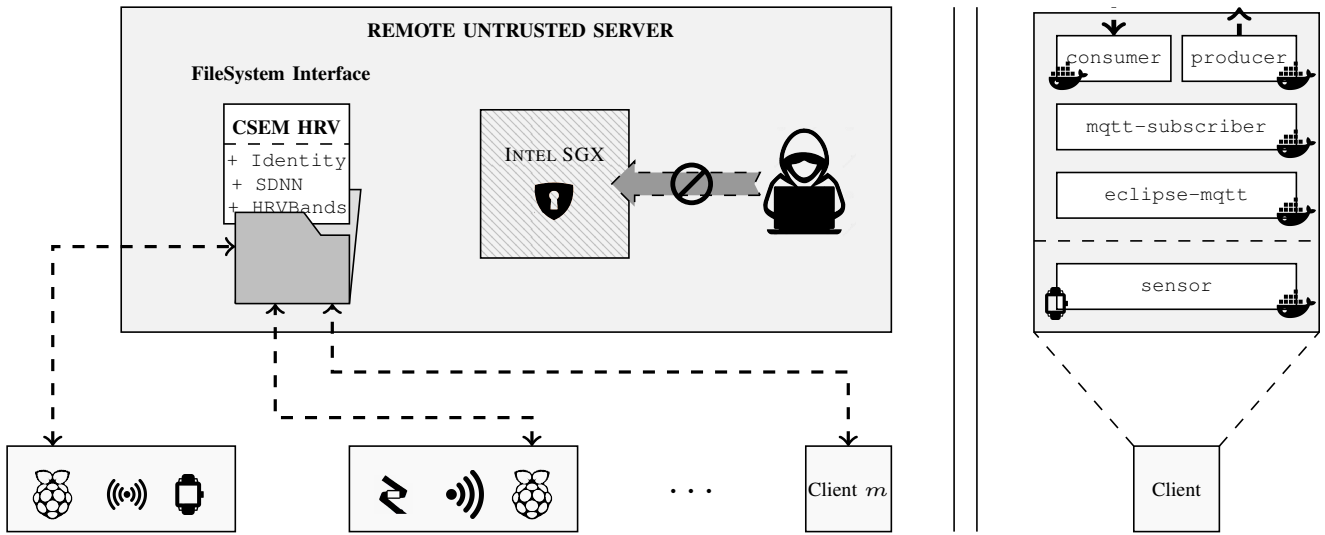
\begin{figure*}[t!]
    \centering
    \input{resources/figs/system-architecture.tex}
    \caption{Schematic representation of the proposed architecture. (Left) Client-server workflow. An arbitrary number of clients composed of a sensor and a gateway communicate with a remote untrusted server with \textsc{Intel SGX} via a FileSystem Interface. The adversarial model assumes a privileged attacker inside the machine. (Right) Client package breakdown. Each client is, in practice, made up of five different components: a \texttt{producer} and \texttt{consumer} service that interact with the remote end, a \texttt{eclipse-mqtt} message broker to distribute the newly generated samples and a \texttt{mqtt-subscriber} to process them, and lastly a \texttt{sensor} that can either be a separate piece of hardware or an artificial \textsc{Docker} service. \label{fig:architecture}}
\end{figure*}

To illustrate that the solution presented is feasible and ready to be deployed, we apply it to a current industry challenge: the processing of cardiac data in the cloud.
Our use case contemplates a scenario where multiple sensors are monitoring the cardiac activity of different users, see Figure~\ref{fig:architecture}.
There are two main systems used by fitness enthusiasts for monitoring heart activity: electrocardiograms (ECG) and photoplethysmograms (PPG).
ECG-based systems measure the heart's electrical activity over time and it is the chosen method by chest-based sensors~\cite{Tamura2018}. PPG-based systems measure the variation of blood volume over time using LEDs and photodiodes. Although less precise, PPGs are the chosen technique by all wrist-based cardiac monitoring sensors~\cite{Parak2015}.

In our case, we focus on the analysis of the Heart Rate Variability (HRV)~\cite{Camm1996}, that is, the analysis of the variation in the time intervals between heartbeats (a.k.a. RR intervals). The HRV is of utmost importance since it has been shown to be a predictor for myocardial infarction~\cite{Kleiger1987,Bigger1992}. We assume that the extraction of these values is performed by the sensor. Figure~\ref{fig:ecg-hrv} depicts a detailed ECG and the information streamed from the sensor.

%% file: resources/figs/ecg-hrv.tex
\resizebox{\linewidth}{!}{
\begin{tikzpicture}
    \pgfdeclaredecoration{single pulse}{initial}{
    \state{initial}[width=\pgfdecoratedinputsegmentlength]
    {%
        \pgfpathlineto{\pgfpoint{0.1*\pgfdecoratedinputsegmentlength}{0mm}}%
        \pgfpathsine{\pgfpoint{0.2\pgfdecorationsegmentlength}{0.15\pgfdecorationsegmentamplitude}}%
        \pgfpathcosine{\pgfpoint{0.2\pgfdecorationsegmentlength}{-0.15\pgfdecorationsegmentamplitude}}%
        \pgfpathlineto{\pgfpoint{0.6\pgfdecorationsegmentamplitude}{0mm}}%
        \pgfpathsine{\pgfpoint{0.1\pgfdecorationsegmentlength}{-0.15\pgfdecorationsegmentamplitude}}
        \pgfpathcosine{\pgfpoint{0.01\pgfdecorationsegmentlength}{0.15\pgfdecorationsegmentamplitude}}%
        \pgfpathsine{\pgfpoint{0.15\pgfdecorationsegmentlength}{\pgfdecorationsegmentamplitude}}%
        \pgfpathcosine{\pgfpoint{0.15\pgfdecorationsegmentlength}{-\pgfdecorationsegmentamplitude}}%
        \pgfpathsine{\pgfpoint{0.15\pgfdecorationsegmentlength}{-0.5\pgfdecorationsegmentamplitude}}
        \pgfpathcosine{\pgfpoint{0.15\pgfdecorationsegmentlength}{0.5\pgfdecorationsegmentamplitude}}%
        \pgfpathlineto{\pgfpoint{1.25\pgfdecorationsegmentamplitude}{0mm}}%
        \pgfpathsine{\pgfpoint{0.8\pgfdecorationsegmentlength}{0.3\pgfdecorationsegmentamplitude}}%
        \pgfpathcosine{\pgfpoint{0.8\pgfdecorationsegmentlength}{-0.3\pgfdecorationsegmentamplitude}}%
        \pgfpathlineto{\pgfpointdecoratedinputsegmentlast}%
    }
    \state{final}{}%
    }

    \fill[gray!10, draw=black] (-0.75, -0.75) rectangle (4.75, 1.5);
    \node at (-0.5, -0.5) {\includegraphics[width=10pt]{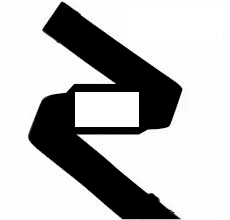}};

    \draw[->, dashed] (4.05,0) -- (4.5,0) node[pos=.5, anchor=north] {\tiny{Time ($s$)}};
    \draw[->, dashed] (-0.5,0) -- (-0.5,1.25) node[anchor=west] {\tiny{Amplitude ($mV$)}};
    \draw[decoration={single pulse,amplitude=8mm,segment length=2mm},decorate] (-0.5,0) -- (1,0);
    \draw[decoration={single pulse,amplitude=8mm,segment length=2mm},decorate] (1,0) -- (2.5,0);
    \draw[decoration={single pulse,amplitude=6mm,segment length=2mm},decorate] (2.5,0) -- (4,0);
    \draw[<->, dashed, thick] (0.10, 0.7) -- (1.5, 0.7);
    \draw[<->, dashed, thick] (1.60, 0.55) -- (2.85, 0.55);
    \node[align=center] at (0, 0.95) {\tiny{$\text{R}_0$}};
    \node[align=center] at (1.2, 0.3) {\tiny{$\text{P}_1$}};
    \node[align=center] at (1.45, -0.2) {\tiny{$\text{Q}_1$}};
    \node[align=center] at (1.7, -0.55) {\tiny{$\text{S}_1$}};
    \node[align=center] at (2.15, 0.38) {\tiny{$\text{T}_1$}};
    \node[align=center] at (1.5, 0.95) {\tiny{$\text{R}_1$}};
    \node[align=center] at (2.9, 0.8) {\tiny{$\text{R}_2$}};

    \node at (5.15, 0) {\includegraphics[width=15pt]{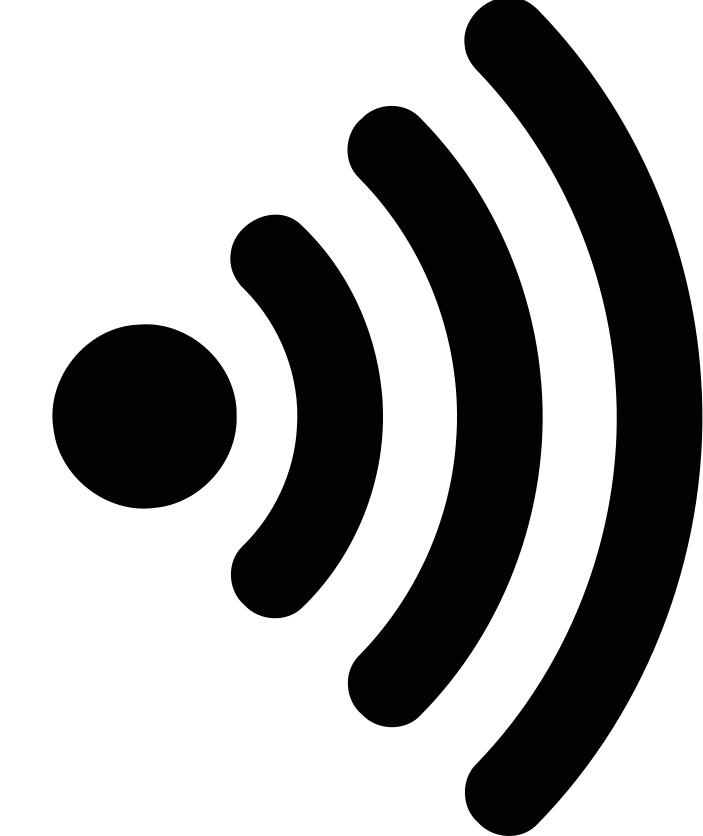}};

    \node[align=center] at (6.65, 1.3) {\tiny{\textbf{\texttt{gateway://data/rr.csv}}}};
    \draw (5.60, -0.75) -- (5.6, 1.1) -- (7.75, 1.1) -- (7.75, -0.4) to[out=180, in=0] (5.6, -0.75);
    \node[align=center] at (6.65, 0.3) {\tiny{\texttt{t}$\left(\text{R}_1\right)$, \xspace\texttt{t}$\left(\text{R}_1\right)$ - \texttt{t}$\left(\text{R}_0\right)$} \\ \tiny{\texttt{t}$\left(\text{R}_2\right)$, \xspace\texttt{t}$\left(\text{R}_2\right)$ - \texttt{t}$\left(\text{R}_1\right)$} \\ \tiny{\textbf{$\vdots$}}};
\end{tikzpicture}}

%% file: resources/figs/system-architecture.tex
\resizebox{\linewidth}{!}{
\begin{tikzpicture}
    \definecolor{ashgrey}{rgb}{0.7, 0.75, 0.71}
    \definecolor{x11gray}{rgb}{0.75, 0.75, 0.75}
    \definecolor{metal}{rgb}{0.43, 0.5, 0.5}

    \fill[gray!10, draw=black] (0, 2.75) rectangle (6.75, 5.6);

    \draw[fill=gray!5] (-1,1) rectangle (1, 1.75);
    \node at (0.6, 1.325) {\includegraphics[width=10pt]{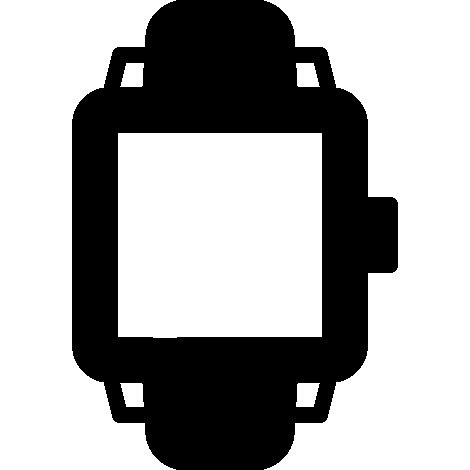}};
    \node at (0.05, 1.325) {\includegraphics[width=10pt]{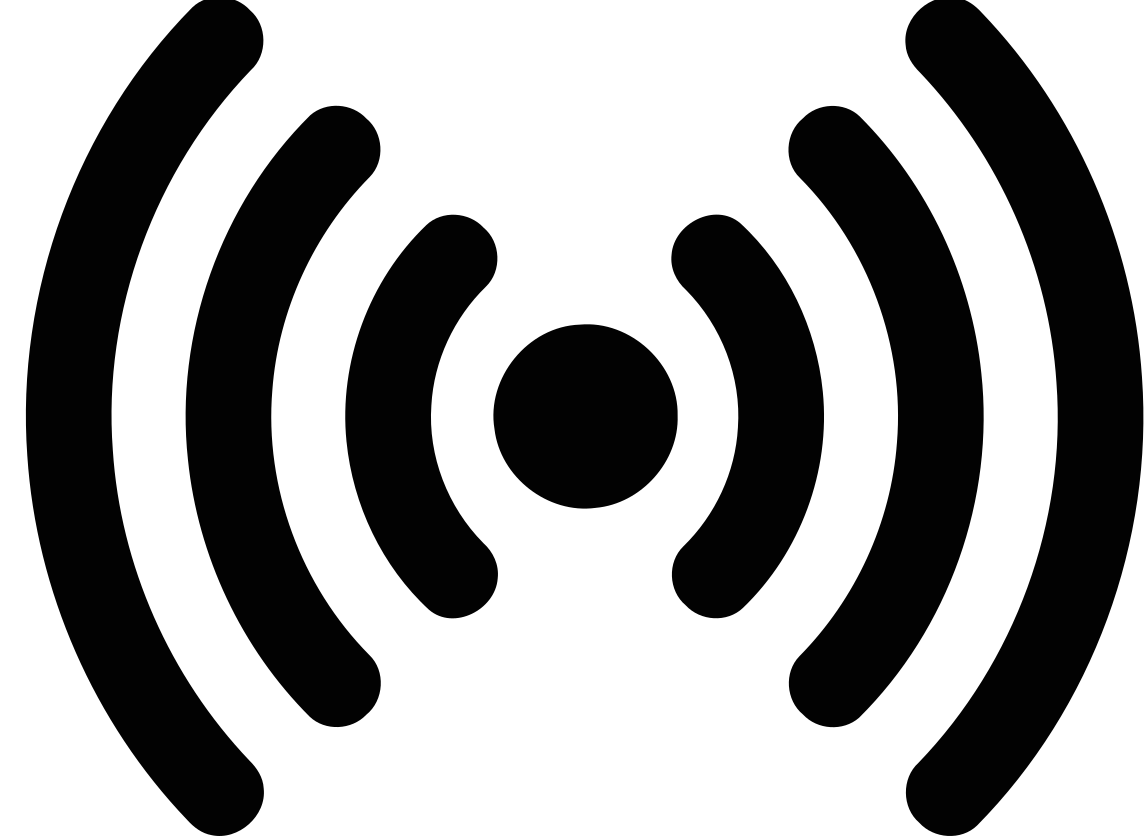}};
    \node at (-0.6, 1.325) {\includegraphics[width=15pt]{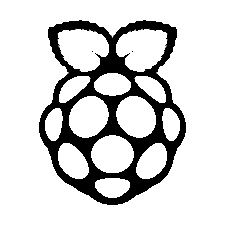}};
    \draw[fill=gray!5] (2,1) rectangle (4, 1.75);
    \node at (2.4, 1.325) {\includegraphics[width=10pt]{resources/img/hrband.png}};
    \node at (3, 1.325) {\includegraphics[width=10pt]{resources/img/wifi-right.png}};
    \node at (3.6, 1.325) {\includegraphics[width=15pt]{resources/img/raspberry.png}};
    \node at (5.025, 1.35) {$\cdots$};
    \draw[fill=gray!5] (6,1) rectangle (6.75, 1.75) node[pos=.5] {\tiny{Client $m$}};
    \draw[fill=gray!5] (8.875,1) rectangle (9.625, 1.75) node[pos=.5] {\tiny{Client}};
    \draw[<->, dashed, thick] (-0.6, 1.75) -- (-0.6, 3.5) -- (1, 3.5);
    \draw[<->, dashed, thick] (3.6, 1.75) -- (3.6, 2) -- (1.35, 2) -- (1.35, 3);
    \draw[<->, dashed, thick] (6.3725, 1.75) -- (6.3725, 2.4) -- (1.85, 2.4) -- (1.85, 3);
    \draw[dashed, thin] (8.875, 1.75) -- (8, 2.75);
    \draw[dashed, thin] (9.625, 1.75) -- (10.5, 2.75);

    \draw[fill=x11gray!50] (1.0, 3) -- (1.1, 3.9) -- (1.6, 3.9) -- (1.75, 4.0) -- (2.1, 4.0) -- (2.0, 3);
    \node at (1.4, 5.0) {\text{\tiny{\textbf{FileSystem Interface}}}};
    \node at (3.5, 5.4) {\text{\textbf{\tiny{REMOTE UNTRUSTED SERVER}}}};
    \draw[fill=white, pattern=north west lines,pattern color=gray!50] (3.15, 3.4) rectangle (4.45, 4.7) node[pos=.5, align=center] {\tiny{\textsc{Intel SGX}} \\[3pt] \includegraphics[width=10pt]{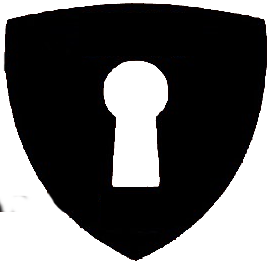}};
    \node at (6, 4.05) {\includegraphics[width=25pt]{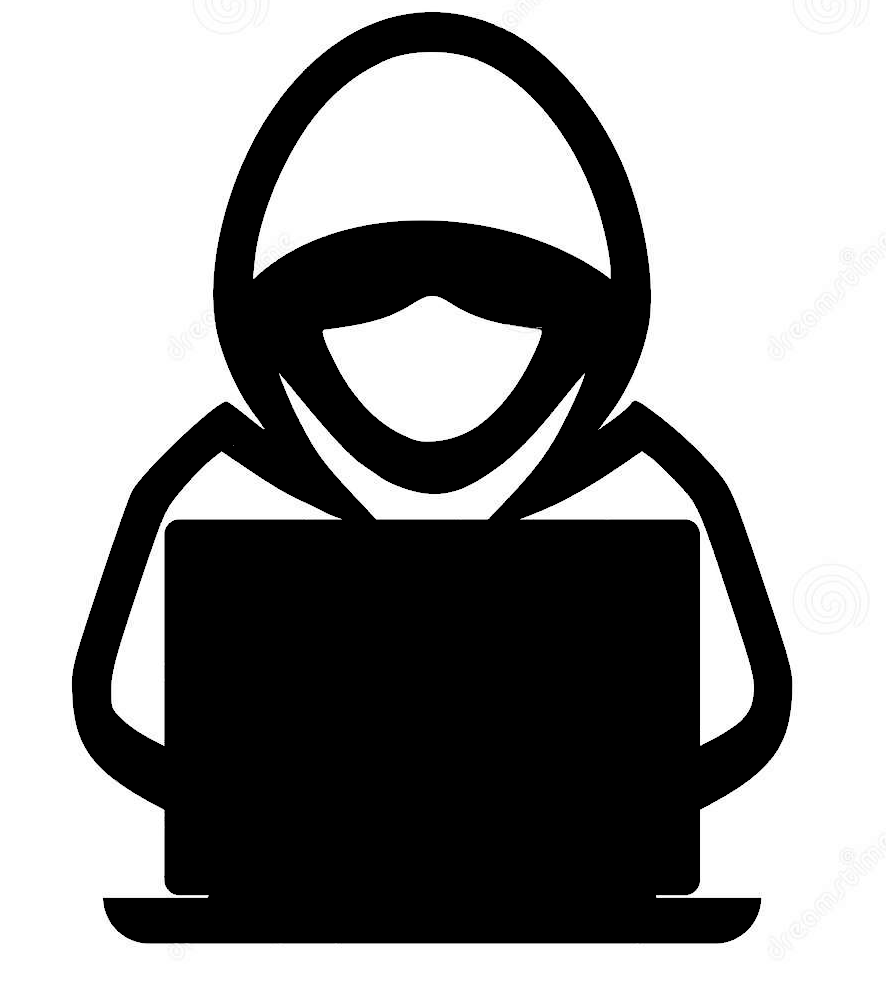}};
    \draw[fill=gray!80, dashed] (5.6, 3.95) -- (4.5, 3.95) -- (4.5, 3.85) -- (4.3, 4.05) -- (4.5, 4.25) -- (4.5, 4.15) -- (5.6, 4.15) -- (5.55, 4.05) -- (5.6, 3.95);
    \node[forbidden sign, draw=black, fill=none, very thick] at (5, 4.05) {};

    \draw[fill=white] (0.9, 3.7) rectangle (2.0, 4.75);
    \node at (1.45, 4.55) {\textbf{\tiny{CSEM HRV}}};
    \draw[dashed] (0.9, 4.4) -- (2.0, 4.4);
    \node at (1.45, 4.25) {\texttt{\tiny{+ Identity}}};
    \node at (1.28, 4.05) {\texttt{\tiny{+ SDNN}}};
    \node at (1.46, 3.85) {\texttt{\tiny{+ HRVBands}}};
    \draw[fill=x11gray] (1.0, 3) -- (1.0, 3.8) -- (1.6, 3.8) -- (1.75, 3.6) -- (2.0, 3.6) -- (2.0, 3) -- (1.0, 3);

    \draw (7.3, 1) -- (7.3, 5.6);
    \draw (7.5, 1) -- (7.5, 5.6);
    \draw[fill=gray!10] (8, 2.75) rectangle (10.5, 5.55);
    \draw[dashed] (8, 3.55) -- (10.5, 3.55);
    \draw[fill=white] (8.2, 2.95) rectangle (10.3, 3.35) node[pos=.5] {\tiny{\texttt{sensor}}};
    \node at (10.3, 3.00) {\includegraphics[width=10pt]{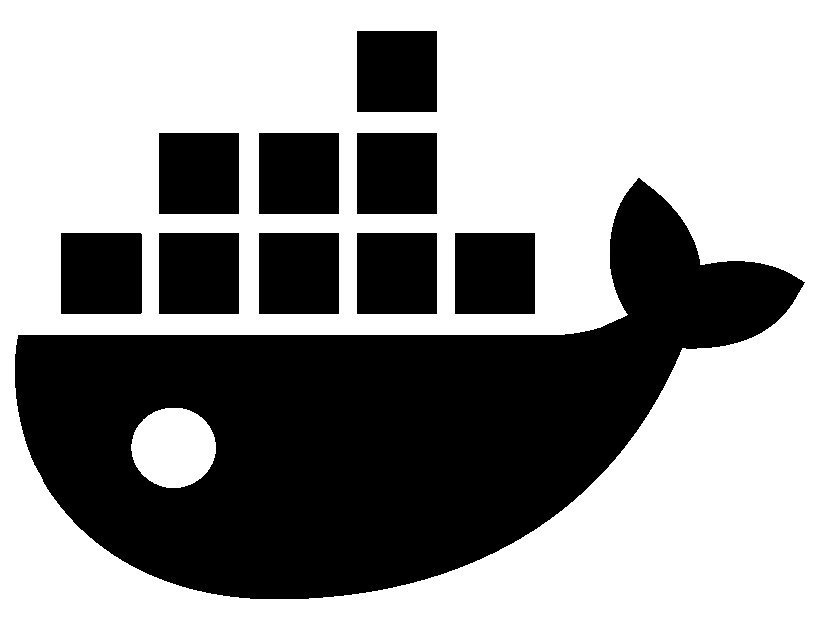}};
    \node at (8.2, 3.00) {\includegraphics[width=8pt]{resources/img/smartwatch.png}};
    \draw[fill=white] (8.2, 3.75) rectangle (10.3, 4.15) node[pos=.5] {\tiny{\texttt{eclipse-mqtt}}};
    \node at (10.3, 3.80) {\includegraphics[width=10pt]{resources/img/docker.png}};
    \draw[fill=white] (8.2, 4.35) rectangle (10.3, 4.75) node[pos=.5] {\tiny{\texttt{mqtt-subscriber}}};
    \node at (10.3, 4.40) {\includegraphics[width=10pt]{resources/img/docker.png}};
    \draw[fill=white] (8.2, 4.95) rectangle (9.2, 5.35) node[pos=.5] {\tiny{\texttt{consumer}}};
    \node at (8.3, 5.00) {\includegraphics[width=10pt]{resources/img/docker.png}};
    \draw[fill=white] (9.3, 4.95) rectangle (10.3, 5.35) node[pos=.5] {\tiny{\texttt{producer}}};
    \node at (10.3, 5.00) {\includegraphics[width=10pt]{resources/img/docker.png}};
    \draw[->, very thick, dashed] (9.8, 5.35) -- (9.8, 5.6);
    \draw[<-, very thick, dashed] (8.7, 5.35) -- (8.7, 5.6);

\end{tikzpicture}}

%% file: sections/architecture.tex
\section{ARCHITECTURE} \label{sec:architecture}

\subsection{System Description}

The proposed architecture follows a client-server scheme.
As depicted in Figure~\ref{fig:architecture}, the system is composed of a remote server located in the cluster (or cloud) with access to \textsc{Intel\textregistered\xspace SGX} and a set of clients distributed among different locations.
In our use-case, each client has two components: a sensor that monitors cardiac data from a user and streams the RR intervals together with their timestamps, and a gateway that aggregates samples and interacts with the remote end.
We impose no restrictions on the type of sensor and, our current gateway implementation could run, for instance, on a \textsc{Raspberry Pi} or a SmartPhone.

On the server, we deployed a modified version of \textsc{Apache Spark} that exploits SGX called \textsc{SGX-Spark}. \textsc{Spark}~\cite{Zaharia2010} is a cluster-computing framework used to develop distributed applications. We implemented the HRV algorithms in Spark's binding for \textsc{Scala}. In particular, we implemented the SDNN temporal analysis and the spectral analysis (HF, LF, VLF)~\cite{Shaffer2017}.

\textsc{SGX-Spark}~\cite{Kelbert2017} is a framework that wraps \textsc{Scala}'s compiler and enables unmodified Spark applications to run critical parts of the processing inside enclaves. In particular, RR intervals are always processed inside Intel SGX.

The client package is divided in four main components: (1) a sensor that streams samples at an approximate rate of 1 to 3~Hz (60 to 180 beats per minute); (2) a \textsc{mqtt}~\footnote{\url{https://mqtt.org/}} message broker; (3) a subscriber on the gateway that processes new samples; and (4) a producer and consumer that interact with the remote end. The link between the gateway and the server is established over \textsc{SFTP} and between 230 and 690~bytes are transferred per second. In our implementation, we are using Eclipse Mosquitto~\footnote{\url{https://mosquitto.org/}} as open source MQTT broker.

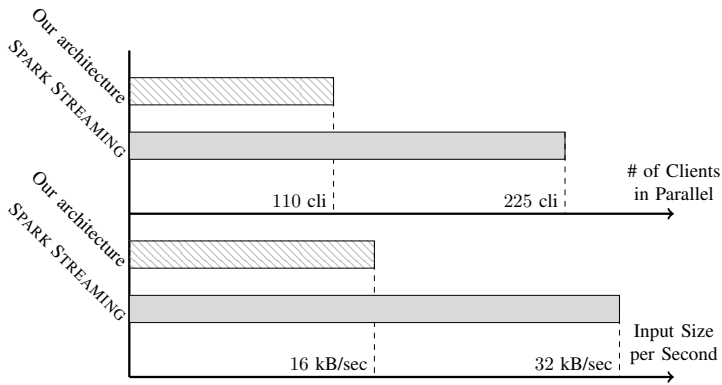
\begin{figure*}[t!]
    \centering
    \input{resources/figs/result-comparison.tex}\hfill
    \begin{tabular}{lcc}
        \toprule
        \textbf{} & \textbf{Proposed} & \textbf{Spark Streaming} \\ \midrule
        \multicolumn{3}{c}{Maximum Number of Clients Served in Parallel}  \\[2pt] \hline
        \# Clients& 110 & 225 \\
        Proportion & 0.49 & 1 \\[1pt] \hline
         &  &  \\[2pt] \hline
        \multicolumn{3}{c}{Maximum Input Load per Second}  \\[2pt] \hline
        Load (kB/sec) & 16 & 32 \\
        Proportion & 0.5 & 1 \\ \hline
         & & \\ \bottomrule
   \end{tabular}
    \caption{Comparison of the maximum number of clients served in parallel and the maximum input load between the proposed architecture and \textsc{Spark Streaming}. (Left) Proportional bar representation of the values at which each system becomes unstable: it can not process the input received during the past 10 seconds in under 10 seconds. (Right) Summary table of the same values. \label{fig:result-comparison}}
\end{figure*}
 
\subsection{Threat Model and Known Vulnerabilities}

Since the communication between the gateway and the server is kept protected by encrypting the data and using a secure transfer protocol (SFTP), our threat model is equivalent to SGX's. We assumes the system software to be completely untrusted (this includes privileged software such as operating systems, BIOS or virtual machine monitors)~\cite{McKeen2013}. To verify the integrity of the code stored in the server, and that the code is really running in an enclave, Intel provides a remote attestation protocol. Once attested, the enclave can be trusted. \textsc{Intel\textregistered\xspace SGX} (in particular the memory encryption engine, MEE~\cite{Gueron2016}) is not designed to be an oblivious RAM, this is, it does not hide memory access patterns. Thus, an adversary could perform traffic analysis attacks.  Other vulnerabilities could be exploited using side-channel~\cite{Schwarz2017} and speculative execution attacks~\cite{VanBulck2018,Weisse2018}. The proposed architecture assumes the client package to be completely trusted. Protecting it lies out of the scope of the project. However, an approach using \textsc{ARM TrustZone} and \textsc{Op-TEE}~\footnote{\url{https://www.op-tee.org/}} could provide similar security guarantees.

%% file: resources/figs/result-comparison.tex
\resizebox{.55\linewidth}{!}{
\begin{tikzpicture}[baseline=(X.base)]
    \draw[very thick] (0, -3) -- (0, 3);
    \draw[very thick, ->] (0,0) -- (10, 0) node[pos=1, rotate=0, align=center, anchor=south, yshift=3pt] {\# of Clients \\ in Parallel};
    \draw[very thick, ->] (0,-3) -- (10, -3) node[pos=1, rotate=0, align=center, anchor=south, yshift=3pt] {Input Size \\ per Second};
    \node at (0, 0) (X) {};

    \node[rotate=-45, anchor=east] at (0, 2) {Our architecture};
    \draw[pattern=north west lines,pattern color=gray!50] (0, 2) rectangle (3.75, 2.5);
    \draw[dashed] (3.75, 2.5) -- (3.75, 0) node[pos=1, anchor=south east] {$110$ cli};
    \node[rotate=-45, anchor=east] at (0, 1) {\textsc{Spark Streaming}};
    \draw[fill=gray!30] (0, 1) rectangle (8, 1.5);
    \draw[dashed] (8, 1.5) -- (8, 0) node[pos=1, anchor=south east] {$225$ cli};

    \node[rotate=-45, anchor=east] at (0, -1) {Our architecture};
    \draw[pattern=north west lines,pattern color=gray!50] (0, -1) rectangle (4.5, -0.5);
    \draw[dashed] (4.5, -0.5) -- (4.5, -3) node[pos=1, anchor=south east] {$16$ kB/sec};
    \node[rotate=-45, anchor=east] at (0, -2) {\textsc{Spark Streaming}};
    \draw[fill=gray!30] (0, -2) rectangle (9, -1.5);
    \draw[dashed] (9, -1.5) -- (9, -3) node[pos=1, anchor=south east] {$32$ kB/sec};

\end{tikzpicture}}

%% file: sections/materials.tex
\section{MATERIALS} \label{sec:materials}

The proposed architecture has two main components: a server side equipped with \textsc{Intel SGX}, and a client package with a sensor and a gateway. The system's assessment is also twofold. Firstly, we run it with data captured from real users to prove that it works in a real case scenario. Secondly, we perform stress tests with artificially generated data and evaluate the overhead introduced by privacy-preserving execution. We are interested in showing that the proposed architecture is also competitive in other scenarios with more stressful workloads.

\subsection{Hardware}
On the server side we use an Intel\textregistered\xspace Xeon\textregistered\xspace CPU E3-1270 v6 @ $3.80$~GHz with 8 cores, 64 GiB RAM, and enclave mode enabled. It is based on \textsc{Ubuntu} $16.04$ LTS (kernel 4.19.0-41900-generic) to support Intel SGX Driver $2.0$. For \textsc{SGX-Spark} we use a development version provided by the LSDS group.

\subsection{Sensors and data}
The used database is obtained from CSEM's proprietary wrist located sensors and chest-located dry electrodes~\cite{Chetelat2015}. In particular, cardiac data is obtained from eight healthy males following a standardized protocol in which they perform a range of physical activities from sedentary to vigorous~\cite{Delgado-Gonzalo2014}. We also augmented the database with simulated data with equivalent statistical moments than the the latter, adapting to the demands in workload of the evaluation stress test.

%% file: sections/evaluation.tex
\section{EVALUATION} \label{sec:evaluation}

To evaluate the system we perform a set of different experiments to assess particular characteristics of the proposed architecture and how do they compare with a normal execution of \textsc{Spark Streaming}, \textit{i.e.} without \textsc{Intel SGX}. We are specially interested in measuring the overhead introduced by privacy-preserving computations. This is, how do enclaves affect the system's latency, or delay, and how it compares in terms of clients that the platform can serve simultaneously.

Each experiment consists of $20$~minutes of stream processing, generating a result sample, or batch, every $10$~seconds. We are interested in measuring the average batch processing time and the memory footprint, that is, how many resources is the job consuming. We consider that a configuration is unstable if the average processing time exceeds $10$~seconds. 

This is, the streaming platform can not process the amount of data received during the last 10 seconds within the next ten seconds.
We simulate data overload via two mechanisms: a single client sending a lot of information per second and many clients being processed simultaneously.
Figure~\ref{fig:result-comparison} summarizes the results obtained for our architecture and \textsc{Spark Streaming}.
In both simulations, Spark is configured to run one master (or driver) process and one worker process with 2~GB of allocated memory.

For the first set of experiments, we deploy a variable number of clients, each streaming samples at approximately $1$~Hz, and measure at which number each configuration becomes unstable.
The system is able to handle $110$~clients simultaneously whilst \textsc{Spark Streaming} is able to handle $225$ users in parallel.

The second set of experiments is performed with a single client, increasing the samples per second the client sends until each system becomes unstable. 
The system is able to process up to $16$~kB per second whilst \textsc{Spark Streaming} manages workloads of up to $32$~kB per second.

To put it in a nutshell, the system seamlessly performs computations on untrusted clouds without compromising user's data with the constraint of approximately halving the system performance. This result is, per se, competitive with other privacy-preserving computing frameworks~\cite{Zheng2017} but is a major improvement, in fact a novelty, for privacy-preserving computing frameworks since it is transparent to the programmer of the final application.

%% file: sections/conclusion.tex
\section{CONCLUSION} \label{sec:conclusion}

In this paper, we have presented a proof of concept of a streaming platform that grants executions on remote, untrusted, servers or clouds with data and code confidentiality and integrity. It provides end-to-end protection transparently to the developer since it runs unmodified \textsc{Apache Spark} applications inside \textsc{Intel SGX}'s enclaves.

We have quantified the impact on overall system performance when protecting health sensitive data from an untrusted cloud provider. More precisely, when performing an HRV analysis, it halves the maximum supported workload and the maximum number of clients the system can process simultaneously. We consider the system to be mature enough to be introduced in a production environment, since it complies with current data protection regulations whilst still maintaining a reasonable performance.

A major further step in reducing the Trusted Computing Base of the overall system would be providing additional protection to the client package. Given the success of enclaves, we suggest considering trusted hardware designed for smaller embedded devices such as \textsc{ARM TrustZone} in combination with a TEE-TEE point to point secure transport link such as \textsc{TaLoS}~\cite{Aublin2017}.